\documentclass[12pt, reprint, superscriptaddress, showpacs, aps, prl]{revtex4-1}
\usepackage{graphicx}
\usepackage{epstopdf}

\def\lsim{\mathrel{\rlap{\lower4pt\hbox{\hskip1pt$\sim$}}
    \raise1pt\hbox{$<$}}}                % less than or approx. symbol
\def\gsim{\mathrel{\rlap{\lower4pt\hbox{\hskip1pt$\sim$}}
    \raise1pt\hbox{$>$}}}                % greater than or approx. symbol

\makeatletter

\newenvironment{Figure}{%
   \par\addvspace{12pt plus2pt}%
   \def\@captype{figure}%
}{%

   \par\addvspace{12pt plus2pt}%

}%

\long\def\@makecaption#1#2{%
  \vskip\abovecaptionskip
  \sbox\@tempboxa{#1: #2}%
  \ifdim \wd\@tempboxa >\hsize
    #1: #2\par
  \else
    \global \@minipagefalse
    \hb@xt@\hsize{\hfil\box\@tempboxa\hfil}%
  \fi
  \vskip\belowcaptionskip}

\makeatother 

\begin{document} 

\title{Measuring the spatial extent of individual localized photonic states} 

\author{Marko Spasenovi\'c}
\affiliation{Center for Nanophotonics, FOM Institute for Atomic and Molecular Physics (AMOLF), Science Park 104, 1098 XG Amsterdam, The Netherlands}
\author{Daryl M. Beggs}
\affiliation{Center for Nanophotonics, FOM Institute for Atomic and Molecular Physics (AMOLF), Science Park 104, 1098 XG Amsterdam, The Netherlands} 
\author{Philippe Lalanne}
\affiliation{Laboratoire Charles Fabry de l'Institut d'Optique, CNRS, Univ Paris-Sud, Campus Polytechnique, RD128, 91127 Palaiseau cedex, France.}
\author{Thomas F. Krauss}
\affiliation{School of Physics and Astronomy, University of St Andrews, North Haugh, Fife, KY16 9SS, UK. }
\author{L. Kuipers}
\affiliation{Center for Nanophotonics, FOM Institute for Atomic and Molecular Physics (AMOLF), Science Park 104, 1098 XG Amsterdam, The Netherlands} 

%\pacs{42.25.Dd, 78.67.Pt}

\begin{abstract} 
We measure the spatial extent of individual localized photonic states in a slow-light photonic crystal waveguide. The size of the states is measured by perturbing each state individually through a local electromagnetic interaction with a near-field probe. We find localized states which are not observed in transmission and show that these states are shorter than the waveguide. We also directly obtain near-field measurements of the participation ratio, from which the size of the states can be derived, in quantitative agreement with the size measured with the perturbation method.
\end{abstract}

\maketitle

%%%%%%%%%%%%%%%%% END OF PREAMBLE %%%%%%%%%%%%%%%%
Waves in disordered media can undergo multiple scattering, resulting in the formation of Anderson-localized states with an associated impeded wave transport. Anderson localization is a universal wave phenomenon, with manifestations in electron transport \cite{Anderson1958}, sound \cite{Hu2008}, matter waves \cite{Billy2008, Roati2008} and light \cite{Wiersma1997}. The spatial extent of localized states, or localization length, is of primary importance. For example, in 1D systems, when this length is larger than the length of the sample, disorder has little effect on wave propagation \cite{Mazoyer2009}. Conversely, when this length is smaller than the sample length, strongly confined states occur and wave transport is severely disrupted. The localization length is an ensemble averaged quantity, usually measured for a series of localization instances, by averaging over frequency or many realizations of disorder. To date, Anderson localization in optical systems has only been observed in transport \cite{Wiersma1997} as narrow resonances in transmission spectra \cite{Bertolotti2005, Topolancik2007a, Topolancik2007, Mazoyer2010}, or using the modified spontaneous emission rates of embedded emitters \cite{Sapienza2010}, or an increase in out-of-plane scattering \cite{LeThomas2009, Garcia2010, Mazoyer2010}. These methods all share the feature that they measure the ensemble average of an observable. From this averaged quantity, one determines the most likely value for some transport property, in the type of system being studied. However, since disorder is stochastic, one does not know how light in a specific structure at a specific frequency will behave. In a sense, the problem is comparable to molecular spectroscopy prior to the success of single-molecule detection, which enabled the study of individual molecules, rather than ensemble averages only \cite{Moerner1989}. 

Here we measure the spatial extent of individual localized photonic states for a single realization of disorder in a photonic crystal waveguide. To emphasize the difference between the ensemble averaged localization length and the measured length of an individual localized state for a single optical frequency for a single realization of disorder, we will use the symbol $L_{ind}$ for the value that we measure. We measure $L_{ind}$ by perturbing the states locally through the light-matter interaction with a near-field probe. We also determine $L_{ind}$ by measuring the individual inverse participation ratio, obtaining values for $L_{ind}$ that agree with those obtained with the local perturbation method.

The idea that disorder in photonic crystals can be used as a model system for the study of localization is as old as the research field of photonic crystals itself \cite{John1987}. Photonic crystal waveguides have been shown to be a good example of true disordered systems in which Anderson localization occurs \cite{Sapienza2010, Mazoyer2010}. For 1D periodic systems, localization always occurs for any non-zero disorder, a fact well established for electrons in 1D potentials \cite{Anderson1958, Jung2011}. As a model system, we investigate for a range of frequencies (wavelengths) near the band edge, a 1-dimensional photonic crystal waveguide (PhCW), formed by a single row of missing holes from a hexagonal lattice of air holes perforating a thin dielectric membrane, as depicted in fig. 1(a). We introduce a near-field probe into the evanescent field above the waveguide and observe an interaction between the probe and the electromagnetic field which changes the spectrum of the probed field. From the change in the spectrum, we measure $L_{ind}$ of individual photonic states. We also measure $L_{ind}$ with a method based on the inverse participation ratio (IPR). The IPR is a quantity related to $L_{ind}$, often used as a measure of disorder. For example, the ensemble-averaged IPR has been used to measure quantum eigenfunctions in ensembles of metallic grains \cite{Prigodin1998}, disorder in 2-dimensional photonic lattices \cite{Schwartz2007}, and to diagnose malignancy in biological cells \cite{Subramanian2008, Pradhan2010}. The $L_{ind}$ measured with our probe-field interaction method agrees with that measured with the IPR method. We identify states for which $L_{ind}$ is smaller than the length of the waveguide. We show that such states do not contribute to transmission significantly.

Photonic crystal waveguides are a versatile tool for controlling the propagation of light. Geometric parameters (such as the size of the holes or thickness of the membrane) control the dispersion of the waveguide modes, which describes the relation between frequency and wavevector of light in the waveguide. The calculated dispersion of our PhCW is shown in fig. 1(b). Important related quantities are the group velocity of light, given by the slope of the dispersion ($v_g \equiv d\omega/dk$), and the density of photonic states. At wavelengths close to 1567.5$\,$nm, the slope of the dispersion becomes shallow, indicating a decreased group velocity. Conversely, the density of states (DOS) increases until the band-edge at $ka/2\pi=0.5$ is reached and (in theory) it diverges to infinity. In practice, however, disorder puts a lower limit on the group velocity (and an upper limit on the DOS) achievable in real systems \cite{Engelen2008, LeThomas2009, Mazoyer2009, Mazoyer2010}. 

\begin{figure}
\centering \includegraphics[width=86mm]{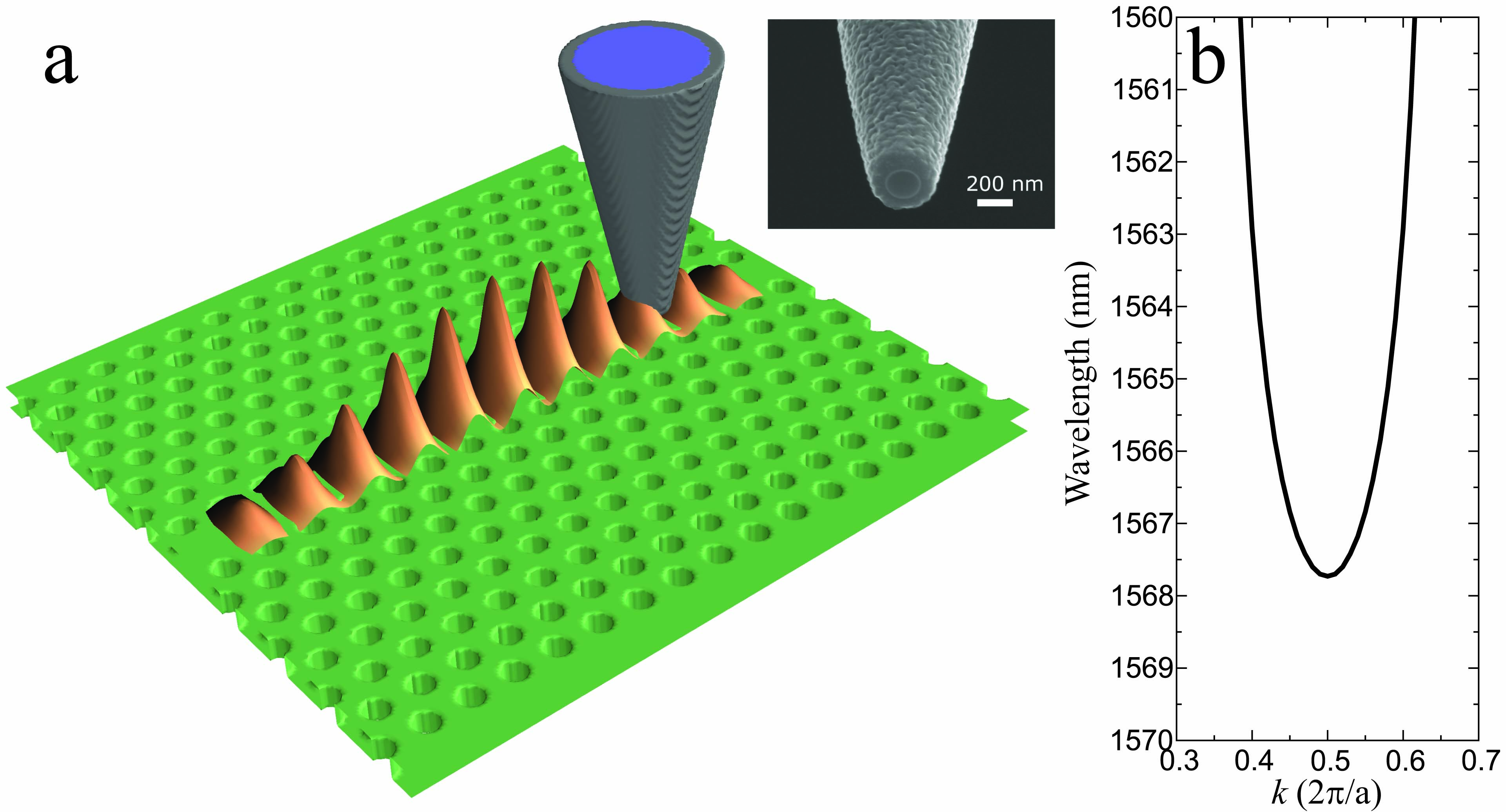}
\caption{{\bf Schematic representation of the sample and near-field probe. a}, An aluminum-coated near-field probe (inset) is scanned over the surface and collects a fraction of the light. The magnetic field of light interacts with the aluminum coating to shift the resonance of localized states. {\bf b}, Calculated dispersion curve for the nominal design.}
\label{f2}
\end{figure}

Although our PhCW is fabricated with state-of-the-art techniques to nominally ideal designs, some residual disorder always remains, resulting in photon scattering. The amount of backscattering into the waveguide scales with the group index squared \cite{Kuramochi2005}, increasing near the band edge. Since the localization length decreases with increasing backscattering \cite{Mazoyer2009}, near the band edge localization length decreases. When the localization length becomes shorter than the waveguide length, $L_{wav}$, photon transport in the waveguide is severely affected. For device lengths typically used in practice ($\approx 100 \,\mu$m), this transition occurs, depending on the fabrication quality, at group indices $n_g \approx 65 $\cite{Mazoyer2010}. 

The resonance behaviour of localized states of light closely resembles that of photonic microcavities \cite{Pendry2008}. In photonic crystals, such microcavities are deliberately engineered defects which confine light to small volumes, with high quality factors and well defined resonance frequencies \cite{Akahane2003}. Anderson localization can be viewed as a formation of such ``cavities'' through the wave interference resulting from multiple scattering by random configurations of disorder in the waveguide. As such, the ``cavities'' occur at unpredictable positions, with unpredictable quality factors and resonance frequencies. We probe the near field of these Anderson localized states using the electrically and magnetically mediated light-matter interactions with a metallodielectric probe.

Our near-field probe is a tapered glass fibre of diameter $\sim 200\,$nm, coated with aluminum of thickness $\sim 150\,$nm (Fig. 1a, inset). Usually, in near-field microscopy, such a probe is introduced into the evanescent field of an optical mode and the electric field of light in photonic eigenstates is detected with high resolution without significantly perturbing the states \cite{Betzig1987}. Above a PhC cavity, however, the probe may perturb the states due to the light-matter interaction between it and the light in the cavity. The electric part of the interaction typically leads to an increase in the effective volume of the cavity, resulting in a red-shift of the resonance, the magnitude of which is inversely proportional to the volume of the photonic state \cite{Koenderink2005}. In the magnetic part of the interaction, the magnetic field of light drives a current through the metal coating, and, in a nanoscopic manifestation of Lenz's law, the magnetic field caused by the current opposes that of the cavity \cite{Burresi2010, Vignolini2010}. The result is that the magnetic interaction shifts the resonance frequency of the cavity to the blue, again with a magnitude inversely proportional to the volume of the state. In our measurements of the PhCW, the magnetic interaction is dominant, like for earlier measurements on a cavity \cite{Burresi2010}, although also the electric interaction plays a role. We insert the probe in the near-field of the localized states and determine the state volume and length from the resultant blue-shift. Hence we measure $L_{ind}$ directly on each instance of localization. %without a need for collecting data on many instances and performing statistical averaging.

Firstly, we measure a transmission spectrum of the unperturbed waveguide, with the probe far from the sample (Fig. 2a). Three spectral regions can be discerned. For a free space wavelength $\lambda \lsim 1561.6\,$nm, the transmission is high, as light is transported through the modes of the waveguide. For $1561.6$\,nm$\lsim \lambda \lsim 1563.2\,$nm the transmission drops, but many large and wide peaks are apparent. Finally for $\lambda \gsim 1563.2\,$nm, the transmission is very low, with a few narrow and sparse peaks appearing, a sign of Anderson localization \cite{Bertolotti2005, Sebbah2006}. 

Next, we bring the near-field probe close to the sample and measure the electric field above the waveguide. The amplitude of the electric field, as a function of laser wavelength and probe position, is depicted in Fig. 2b. Measuring the electric field at the exit of the PhCW (at the position $x$ = 82 $\,\mu$m), one notes that the magnitude of the measured electric field correlates with the measured transmission for all wavelengths. For $\lambda \lsim 1561.6\,$nm (top of the image), we operate far above the band edge, and light propagates in the Bloch modes of the periodic structure. The electric field amplitude at the exit is similar to that at the entrance. For $1561.6$\,nm$\lsim \lambda \lsim 1563.2\,$nm, the magnitude of the measured electric field at the exit decreases on average. However, we also occasionally observe sharp increases in the amplitude in the waveguide, corresponding to the first localization peaks; for example close to $\lambda \approx 1562.0\,$nm. Another example of an extended localization instance is seen at $\lambda = 1563.35\,$nm. Such features seem to be spatially extended over the length of the waveguide. At longer wavelengths ($\lambda \gsim 1563.2\,$nm), populated states become sparse, but we observe light inside the waveguide at wavelengths at which transmission is negligible. At these wavelengths, the light seems to have more intensity close to the entrance of the waveguide than at the exit, indicating states which are shorter than the waveguide and which are populated by the light incident from the input waveguide on the left. It is noteworthy that the periodic amplitude modulation owing to beating between forward- and backward-propagating Bloch modes persists throughout the localization regime, an indication that at all frequencies the light populates modes of the nominal, perfect structure.

\begin{Figure}
\centering \includegraphics[width=86mm]{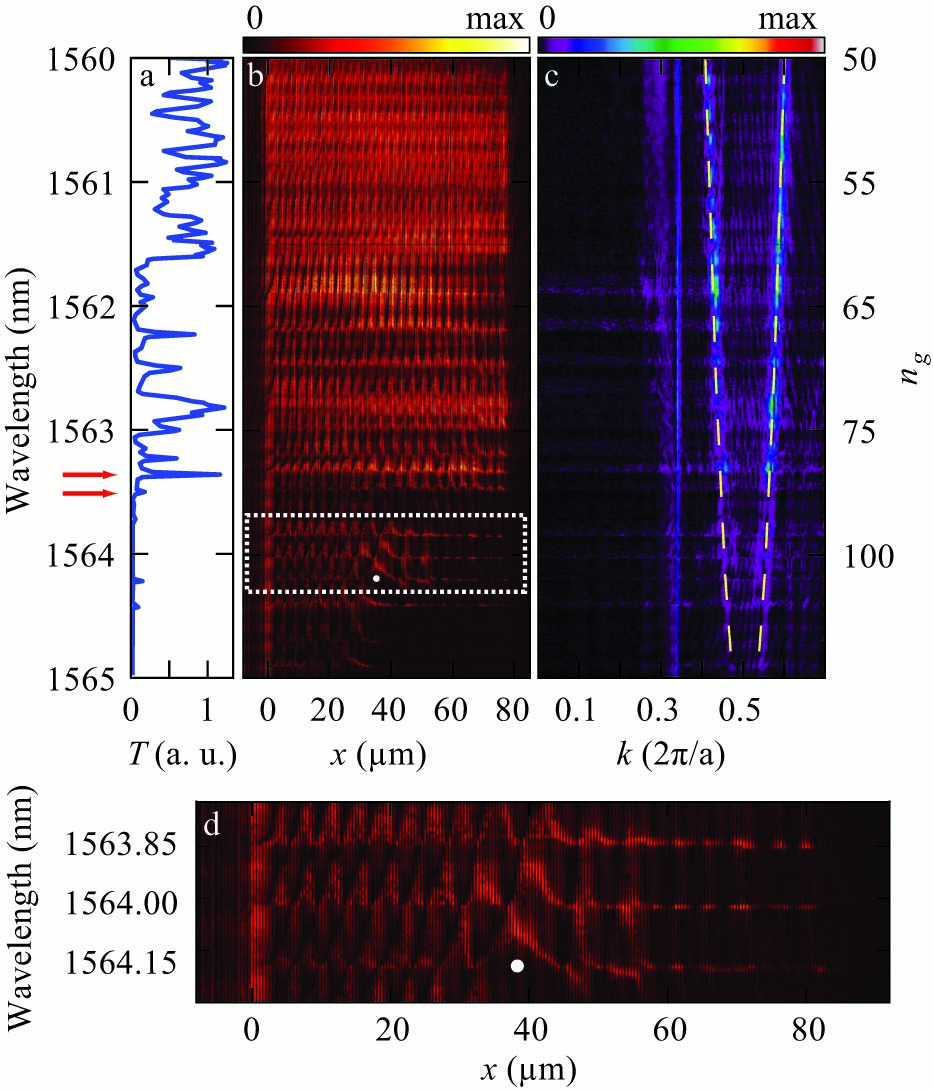}
\caption{{\bf Waveguide transmission, near-field amplitude, and dispersion. a}, Transmission spectrum, $T$, of the PhCW. {\bf b}, Near-field amplitude collected through the probe, as a function of wavelength and probe position along the waveguide. Light enters the PhCW from the left, at position $x$ = 0. {\bf c}, Band structure of the waveguide. Waveguide modes appear as lines in the spectrum. A parabolic fit is depicted by the dashed yellow line. The group index, $n_g$, is obtained by taking the derivative of this line. Localized states appear as horizontal lines in the Fourier spectrum aligning with peaks in transmission. The almost vertical line around $k=0.33$ is the light line in the silicon slab. {\bf d}, Closeup of the near-field region indicated with a dotted white box in part b).}
\end{Figure}

Figure 2(c) depicts the dispersion of the waveguide mode, obtained via a fast Fourier transform of the real-space data to reciprocal space \cite{Engelen2007}. The dispersion curve of the PhC waveguide follows the behavior expected from calculation (Fig. 1b). It is noteworthy that the measured bandstructure follows that of the nominal, perfect structure throughout the localization regime. In the band structure, localized states appear as horizontal lines spanning reciprocal space; the broad distribution of wavevectors results from the high spatial confinement. Away from the band edge (top of the image), the dispersion curve can be approximated by a parabola. A parabolic fit to this part of the curve results in the dashed yellow line. The group index, shown on the right-hand axis, is obtained by taking the first-order derivative of the fit. Theory and measurements involving statistical averaging \cite{Mazoyer2010} have shown that the average length of localization instances becomes of the order of the waveguide length at $n_g \sim 65$. This group index corresponds very well with our first indication of localization, at $\lambda$ = 1562{\,}nm. 

The near-field probe collects the electric field with high resolution, but at the same time we observe its interaction with the electric and magnetic fields as shifts in the spectral position of the localized resonances. This shifting is most evident for long wavelengths ($\lambda > 1563.5${\,}nm), at the bottom of Fig. 2b. We observe that for certain spatial positions of the probe, the maximum of the electric field is shifted from its original spectral position towards shorter wavelengths. One such position is indicated by a white dot in the image. A closeup of the region with the strongest shifts, indicated with a dotted white box in Fig. 2b, is given in Fig. 2d. The shift occurs periodically with spatial position, as a result of interference between forward- and backward-propagating Bloch modes. 

In engineered PhC microcavities, the spectral shift was shown to be given by \cite{Koenderink2005, Burresi2010}:

\begin{eqnarray}\label{e1}
\frac{\Delta \lambda (\bf{r_{\textrm{pr}}})}{\lambda} & \approx & \alpha^{e}_{\textrm{eff}}\left.\frac{|E_{0}|^{2}}{U_{0}}
+ \alpha^{m}_{\textrm{eff}} \frac{|B_{0}|^{2}}{U_{0}}\right.\nonumber \\
& = & \frac{\alpha^{e}_{\textrm{eff}}}{2V_{\textrm{cav}}}\left.\frac{|\bf{E}(\bf{r_{pr}})|^2}{\textrm{max}[\epsilon(\bf{r_{pr}})|\bf{E}|^2]}
+ \frac{\alpha^{m}_{\textrm{eff}}}{2V_{\textrm{cav}}} \frac{|H_z(\bf{r_{pr}})|^2}{\textrm{max}[\mu_{0}|H_z|^2]}\right.\nonumber \\
& = &
\frac{\alpha^{e}_{\textrm{eff}}}{2V_{\textrm{cav}}}\left.\beta_{E}
+\frac{\alpha^{m}_{\textrm{eff}}}{2V_{\textrm{cav}}}\beta_{H},\right.
\end{eqnarray}

\noindent where $U_0$ is the energy of the unperturbed cavity field; $E_0$ and $B_0$ are, respectively, the total electric and magnetic fields of the unperturbed cavity; $\alpha^{e}_{\textrm{eff}}$ is the effective polarisability of the probe; $\alpha^{m}_{\textrm{eff}}$ is the magnetic polarisability of the probe; $V_{\textrm{cav}}$ is the cavity volume (area*length); $\epsilon(\bf{r_{pr}})$ is the electric permittivity at the position of the probe; $\mu_{0}$ is the magnetic constant; max$|E|$ is the maximum magnitude of the electric field in the sample; $|E(r_{pr})|$ is the magnitude of the electric field at the position of the probe; $|H_z(r_{pr})|$ is the magnitude of the component of the magnetic field which points into the probe at the position of the probe, and max$|H_{z}|$ is the maximum magnitude of $H_{z}$ in the sample. 

The electric and magnetic polarisabilities of our probes were calculated by Burresi \textit{et al.} \cite{Burresi2010} to have values of $\alpha^{e}_{\textrm{eff}}=3 ~\textrm{x} ~10^{-21} \textrm{m}^{3}\epsilon_{0}$ and $\alpha^{m}_{\textrm{eff}}=-12 ~\textrm{x} ~10^{-21} \textrm{m}^{3}/\mu_{0}$, respectively. To estimate the other parameters in equation (\ref{e1}), we calculated the eigenmodes of the unperturbed photonic crystal waveguide by using the MIT Photonic Bands Package \cite{Johnson2001}. The eigenmodes at the wavelengths of our study have an effective cross sectional area $A$ of $1.12a^2$, with $a$ the period of the PhC lattice. The ratios substituted by variables $\beta_{E}$ and $\beta_{H}$ at the height of the probe are both equal to 0.22. We are now in a position to estimate $L_{ind}$ as

\begin{equation}\label{e3}
L_{ind}=0.22 \frac{\lambda}{\Delta\lambda}\frac{\alpha_{eff}^{e}+\alpha_{eff}^{m}}{2A},
\end{equation}
where we have used $V_{cav}=L_{ind}A$ and take $A$ to be the same as that of the waveguide mode. Applying eq. 2 to clearly identifiable localized states from fig. 2 results in a determination of their localization length. We depict these measured localization lengths as a function of wavelength and group index by the blue points in fig. 3. To the best of our knowledge, this is the first time localized state length has been measured directly on individual localized states in a photonic system. The localization length generally becomes smaller at higher $n_g$'s, as the backscattering of waves increases \cite{Kuramochi2005} and using our method we can precisely measure that for this particular waveguide Anderson localization sets in, when $L_{ind} < L_{wav}$, at $n_g > 80$, at which point the waveguide is no longer useful for reliable optical transport. At higher values of $n_g$, we observe states in the near field of the waveguide, but they are shorter than the waveguide and do not contribute significantly to transmission.

Using eq. 2 we can determine $L_{ind}$ by measuring the perturbation of localized states via their spectral shift. We can also determine the degree of localization by measuring IPR \cite{Wegner1980}, also in the near field. The IPR is defined as \cite{Wegner1980, Cohen1983}

\begin{equation}\label{e4}
IPR=\int{|\bf{E}(\bf{r})|^{4}d\bf{r}},
\end{equation}
\noindent and is proportional to localization length \cite{Cohen1983, Schreiber1985}. In 1D, and normalised to the length of the waveguide, the IPR takes the form

\begin{equation}\label{e5}
IPR=\frac{L_{loc}}{L_{wav}} = \frac{\int{|\bf{E}(\bf{r})|^{4}d\bf{r}}}{(\int{|\bf{E}(\bf{r})|^2d\bf{r}})^2},
\end{equation}
where $L_{loc}$ is the ensemble averaged localization length. As we have measured the electric field as a function of position and we know $L_{wav}$, we determine $L_{ind}$ using eq. 4, assuming that the equation holds for measurements of individual states as well as ensemble averages. We plot the $L_{ind}$ obtained with this IPR method as green points in Fig. 3, together with $L_{ind}$ obtained by measuring the spectral shift of localized states. The agreement between the two methods is quantitative. On the right side of the plot, where $L_{ind}$ becomes shorter than $L_{wav}$, the two methods give the same results. The localization length decreases towards longer wavelengths (higher group indices), dropping to $\approx L_{wav}/3$ at the lowest energy localized state of this study. As the wavelength is decreased away from the band edge, the localization length rises, and becomes on the order of the waveguide length at $\lambda \approx 1563.5\,$nm ($n_g\approx 80$). At shorter wavelengths, $L_{ind} \approx L_{wav}$, with sharp dips at the positions of localization instances. At $\lambda<1561\,$nm ($n_g<55$), $L_{ind} > L_{wav}$, indicating extended states.

\begin{figure}
\centering \includegraphics[width=86mm]{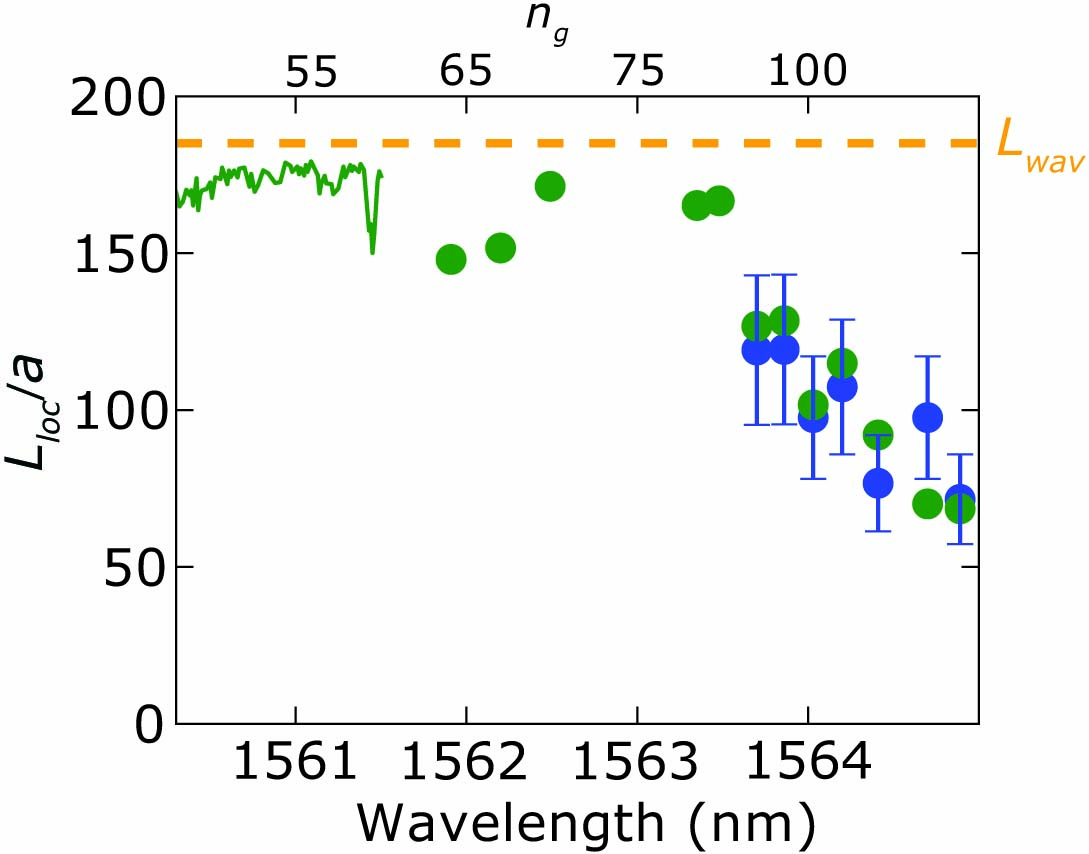}
\caption{{\bf Dispersion of the localization length.} The blue dots are points measured with the near-field interaction method. The green data points are measured with the IPR method. The horizontal dashed line indicates $L_{wav}$, the length of the waveguide.}
\label{f3}
\end{figure}

Taking figures 2 and 3 as a whole, we now conclude that three transport regimes can be identified. For $\lambda \lsim 1561.6\,$nm, $L_{ind}>L_{wav}$ and photon transport in the waveguide is hardly affected by disorder. For higher wavelength the localization length decreases, until for $1561.6\,$nm$\lsim \lambda \lsim 1563.2\,$nm, we have a situation where $L_{ind} \approx L_{wav}$. In this regime photon transport becomes unpredictable, with almost equal probability of high or low transmission at any given wavelength. Finally, for the longest wavelengths ($\lambda \gsim 1563.2\,$nm), the localization length has decreased well below the waveguide length, and photons can only be transmitted in sporadic cases. Populated localized states do exist in this regime, however, but their length is shorter than the length of the waveguide. It is exactly these states which are useful for applications requiring strongly localized fields, such as quantum optics \cite{Sapienza2010}. In all regimes of transport, our method of measuring the $L_{ind}$ from the probe-field interaction gives the same result as the IPR method, which is commonly used to measure the statistically averaged degree of disorder in various systems \cite{Schwartz2007, Subramanian2008, Pradhan2010}.

Our work shows that a key parameter in localization studies can be measured directly for a single realisation of disorder. The $L_{ind}$ of a single instance cannot be measured by observing transmission, and states for which $L_{ind}<L_{wav}$ are often not even observable in transmission. Our measurement approach can be extended to investigate and understand the behavior of any disordered system where electromagnetic waves are present at the surface. The technique can easily be extended to finite size 2D systems, which are far from trivial, and in which the interplay between localization and out-of-plane scattering still has to be fully understood. Notably, we do not need to perform the measurement over the entire length or volume of the state in order to determine $L_{ind}$. Our method could thus be used to determine the volume of localized states just below the surface of a 3D disordered (photonic crystal) structure. From the application point of view, a recent study shows that localization in photonic crystal waveguides can initiate lasing from quantum dots \cite{Yang2011}. The gain of random lasers has been theoretically predicted to depend exponentially on localization length \cite{Burin2002}, but shown numerically to have a power-law dependence on localization length \cite{Li2001}. Our technique provides an ideal method to resolve such dual predictions, by allowing studies of gain and state length for each individual localized state. We further expect our local perturbation method to also work for other wave phenomena, such as acoustics. The fact that we exploit the magnetically induced light-matter interaction also reveals the fascinating opportunity to study disorder in systems with a strong magnetic field at optical frequencies, such as metamaterials \cite{Linden2004, Burresi2009}.\\

We thank Ad Lagendijk, Femius Koenderink, and Sanli Faez for useful discussions. This work is part of the research program of the Stichting voor Fundamenteel Onderzoek der Materie (FOM), which is financially supported by the Nederlandse Organisatie voor Wetenschappelijk Onderzoek (NWO). We thank the EC Marie Curie Scheme (contract MEST-CT-2005-021000).\\

\end{document}